 \documentclass[final,5p,times,twocolumn,authoryear]{elsarticle}

\usepackage{amssymb}
\usepackage{lipsum}
\usepackage{multirow}
\usepackage{xcolor}
\usepackage{color}
\usepackage{adjustbox}

\journal{Astronomy $\&$ Computing}

\begin{document}

\begin{frontmatter}

\title{Co-SOM: Co-training for photometric redshift estimation using Self-Organizing Maps}




\author[label1,label2,label3]{Alvaro Callejas-Tavera}
\affiliation[label1]{organization={Universidad Nacional Autónoma de México, Posgrado en Ciencia e Ingeniería de la Computación},
            addressline={Tablaje Catastral No. 6998}, 
            city={Mérida},
            postcode={97357}, 
            state={Yucatán},
            country={México}}
            
 \ead{callejas.alvaro@aries.iimas.unam.mx}
 
 \author[label2]{Erik Molino-Minero-Re}
 \affiliation[label2]{organization={Universidad Nacional Autónoma de México, Instituto de Investigaciones en Matemáticas Aplicadas y en Sistemas, Unidad Académica Yucatán},
             addressline={Tablaje Catastral N°6998},
             city={Mérida},
             postcode={97357},
             state={Yucatán},
             country={México}}
 \ead{erik.molino@iimas.unam.mx}      
  
 \author[label3]{Octavio Valenzuela}
 \affiliation[label3]{organization={Universidad Nacional Autónoma de México.Instituto de Astronomía.},
             addressline={A.P. 70-264},
             postcode={04510},
             city={Ciudad de Mexico},
             country={Mexico}}
 \ead{octavio@astro.unam.mx}

\begin{abstract}
The upcoming galaxy large-scale surveys, such as the Vera C. Rubin Observatory’s Legacy Survey of Space and Time (LSST), will generate photometry for billions of galaxies. The interpretation of large-scale weak lensing maps, as well as the estimation of galaxy clustering, requires reliable redshifts with high precision for multi-band photometry. However, obtaining spectroscopy for billions of galaxies is impractical and complex; therefore, having a sufficiently large number of galaxies with spectroscopic observations to train supervised algorithms for accurate redshift estimation is a significant challenge and an open research area. We propose a novel methodology called Co-SOM based on Co-training and Self-Organizing Maps (SOM), integrating labeled (sources with spectroscopic redshifts) and unlabeled (sources with photometric observations only) data during the training process, through a selection method based on map topology (connectivity structure of the SOM lattice) to leverage the limited spectroscopy available for photo-z estimation. We utilized the magnitudes and colors of Sloan Digital Sky Survey data release 18 (SDSS-DR18) to analyze and evaluate the performance, varying the proportion of labeled data and adjusting the training parameters. For training sets of 1\% of labeled data ($\approx20,000$ galaxies) we achieved a performance of bias $\Delta z = 0.00007 \pm 0.00022$, precision $\sigma_{zp} =0.00063 \pm 0.00032$ and outlier fraction $out\_frac =0.02083 \pm 0.00027 $. Additionally, we conducted experiments varying the volume of labeled data, and the bias remains below $10^{-3}$, regardless of the size of the spectroscopic or photometric data. These low-redshift results demonstrate the potential of semi-supervised learning to address spectroscopic limitations in future photometric surveys.
\end{abstract}




\end{frontmatter}




\section{Introduction}

In recent years, the deployment of deep neural networks (DNN) for classification and regression tasks (supervised learning) in large databases and the generation of massive datasets have grown substantially \citep{D_Isanto_2018}. Consequently, these methodologies have been widely adopted in multiple scientific domains \citep{Barchi_2020}. Machine learning (ML) has become pivotal in modern astrophysics, enabling breakthroughs in data-intensive challenges across the field. 

In astronomy, there is a strong interest in constructing highly accurate maps of the Universe by estimating redshift values and classifying objects from imaging data \citep{D_Isanto_2018, Sadeh_2016, Carrasco_Kind_2014}. Some widely used learning methodologies for photometric redshift estimation are: GPz \citep{Almosallam_2016}, which is based on Bayesian learning; ANNz2 \citep{Sadeh_2016}, which employs neural networks combined with ensembles of decision trees; and SOMz \citep{Carrasco_Kind_2014}, which relies on self-organizing maps.

In the forthcoming decade, new galaxy large-scale surveys such as LSST, RST, EUCLID \citep{LSST_2019, RST_2023, EUCLID_2024} will deliver multi-band photometry for billions of galaxies. Each galaxy must have a reliable redshift with high precision ($\Delta z < 0.003$) to correct potential systematics that can affect the shear signal \citep{Laureijs_2011, Schmidt_2020, Hoyle_2016, Masters_2015, Hildebrant_2010}. However, the number of spectroscopic redshifts (labeled data)  will not be sufficient ($\approx 1\%$), since obtaining spectroscopy for billions of galaxies is impractical, complex, and time-consuming \citep{Pasquet_2018}. 

Supervised methods need $\approx 10\%$ of labeled data to guarantee robust estimations \citep{Pasquet_2018, Schmidt_2020}. Still, given the limitations of labels, these methods may suffer from bias, their practicality is hindered by the considerable demand for labeled data during the training stage. Nevertheless, this challenge can be mitigated by leveraging galaxy photometry through machine learning methods. Alternative network architectures, such as Self-Organizing Maps (SOMs), are capable of organizing and preserving the topological relationships inherent in the input data without relying heavily on labeled samples, leveraging unsupervised learning \citep{Kohonen_1989, Pedro_2019}.

Consequently, this architecture has gained popularity and has been applied within the field. \cite{Buchs_2019} and \cite{Wright_2025} exploit the data organization properties of SOMs to characterize redshift distributions for weak lensing analyses. Similarly, \cite{Wright_2020} employs SOMs to achieve a high-fidelity discrimination of the color–redshift relation, enabling the calibration of the KiDS+VIKING-450 dataset as presented in \citet{Wright_2019, Hildebrandt_2020}.  Furthermore, \cite{Carrasco_Kind_2014} extensively explores the use of SOMs for estimating galaxy photometric redshift probability density functions (PDFs), introducing the concept of combining the outputs of multiple maps (the SOM Atlas) to generate a weighted final estimate.

The advancement of models that combine labeled and unlabeled data to improve accuracy has gained increasing importance \citep{Jesper_2019}. By integrating unlabeled data into the training process, it has been observed that the training error decreases after only a few iterations. This trend has led to the development and diversification of the Semi-Supervised Learning (SSL) paradigm, positioning it as an intermediate framework between supervised and unsupervised learning \citep{Chapelle_2009, Zhu_2008}.

SSL strategically leverages the limited availability of labeled data by incorporating large amounts of unlabeled data into the training process. However, these techniques require bias-mitigation mechanisms to ensure the reliable integration of unlabeled data, as they are inherently prone to the problem of noisy labels. Two widely adopted strategies to mitigate label noise are predictive confidence and uncertainty modeling \citep{Du_2020, Hu_2020, Wang_2021}.

This work introduces a novel semi-supervised methodology for photometric redshift estimation based on the co-training of Self-Organizing Maps (Co-SOM). Using only $\approx 1\%$ spectroscopic labels, the method produces point estimates that achieve bias values competitive with LSST photo-z targets rather than full posterior PDFs. However, it should be noted that completeness and selection effects were not simulated, as addressing them lies outside the scope of this work. Unlike previous works with SOM \citep{Carrasco_Kind_2014, Wright_2020, Hildebrandt_2020}, Co-SOM integrates unlabeled data into the training process through a selection strategy that exploits the topological relationships among neurons. In addition, bias-mitigation techniques are applied at every stage, and a new training scheme for the SOMs improves accuracy. The simultaneous training of multiple SOMs further increases robustness, precision, and adaptability to variations in photometric data. Extending this framework to construct full $N(z)$ distributions for cosmological analyses is left for future work.

The paper is structured as follows: Section 2 describes the Self-Organizing Map. Section 3 develops the concept of semi-supervised learning and explains the Co-training method. Section 4 presents a discussion of the astronomical data set used for testing. Section 5 introduces the Co-SOM algorithm. Section 6 reports and discusses the results obtained in the different tests. Section 7 summarizes the main results obtained and presents the potential directions for future research.

\section{Self-Organizing Map}

\begin{figure}[h!]
    \centering
    \includegraphics[width=0.3\textwidth]{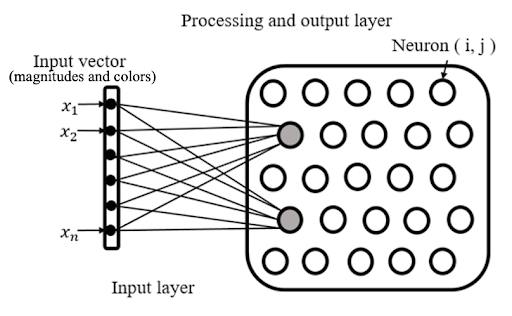}
    \caption{The input layer consists of data inputs (magnitudes and colors), each represented by an $n$-dimensional feature vector. During training, the output layer is functionally equivalent to the processing layer, as both share the same topological structure of the self-organizing map. Each neuron in this layer has an $n$-dimensional weight vector, matching the input data dimensionality, and is assigned a fixed coordinate ($i,j$) on a two-dimensional grid. This topological structure enables the SOM to preserve the spatial relationships inherent in the input space.}
    \label{fig:som}
\end{figure}

The Self-Organizing Map (SOM) is an unsupervised, non-parametric neural network primarily used to generate discrete representations of input data, typically arranged in a map-like structure. This method is advantageous in the exploratory data analysis phase, as it facilitates visualization and investigation of high-dimensional data structures \citep{Kohonen_1989}. During the SOM training process, input vectors are projected onto a lower-dimensional grid composed of units, commonly referred to as neurons. This mapping enables visualization and exploration of the intrinsic properties of the data \citep{Vesanto_2000}. An example of a SOM representation is shown in Figure \ref{fig:som}.

During training, SOM involves a competitive learning process where each input vector is compared to all neurons using a predefined distance metric, typically the Euclidean distance, to identify the Best Matching Unit, which guides the adaptation of the surrounding neurons according to a neighborhood function. SOMs are particularly valued for their ability to preserve the topology of input data via a neighborhood function during training.

The spatial configuration of neurons and the pattern of their interconnections, commonly referred to as the map architecture, directly influences the degree to which the input data topology is preserved. If neighborhood relationships among neurons are not adequately maintained, intrinsic patterns within the data may be distorted or lost. The architecture defines the connectivity between adjacent neurons: in a rectangular architecture, each neuron is typically connected to four immediate neighbors, whereas in a hexagonal architecture, each neuron is connected to six.

\subsection{SOM algorithm}
Let the input data vector with $m$ features (magnitudes and colors) be denoted as $\vec{x}_{n,m}$, where $n$ indexes the data sample. Similarly, let the neuron vector be denoted as $\vec{w}_{k,m}$, where $k$ is the neuron index. Initially, the vectors in the neuron lattice are randomly initialized, and the parameters $\alpha_{0}$, $\delta_{0}$, and $T$ are set. These parameters are defined as follows:

\begin{itemize}
    \item $\alpha_{0}$: The learning-rate factor
    \item $\delta_{0}$: The initial radius of the neighborhood function 
    \item $T$: Number of epochs in the training phase
\end{itemize}

At each step $t$, the Euclidean distances between vectors from the input data and each neuron are calculated according to the following equation (Equation \ref{eq_1}):

\begin{equation}
d_{k}(t) = d(\vec{x}(t),\vec{w_{k}}) = \sqrt{\sum_{i=1}^{m} (x_{i}(t)-w_{k,i}(t))^{2}}
\label{eq_1}
\end{equation}

The Best Matching Unit is denoted by the subscript $b$, as defined in Equation \ref{eq_2}:

\begin{equation}
d_{b}(t) = \min_{k} d_{k}(t)
\label{eq_2}
\end{equation}

The Best Matching Unit (BMU) and its neighboring neurons, as determined by the previously defined architecture and located within the specified neighborhood radius, are updated throughout the training process. This updating mechanism ensures that the input data is accurately represented within the map, preserving its topological structure throughout the training phase, as described in Equation \ref{eq_3}:

\begin{equation}
\vec{w_{k}}(t+1) = \vec{w_{k}(t)} + \alpha(t)H_{b,k}(t)[\vec{x}(t)-\vec{w_{k}(t)}]
\label{eq_3}
\end{equation}

Here, $\alpha(t)$ denotes the learning rate at iteration $t$, and $H_{b,k}(t)$ represents the Gaussian neighborhood function centered at the Best Matching Unit (BMU) $b$ for neuron $k$, as defined in Equation \ref{eq_4}.

\begin{equation}
H_{b,k} = \exp{\frac{-D_{b,k}^{2}}{2*\delta(t)^{2}}}
\label{eq_4}
\end{equation}

This function quantifies the influence of the BMU on its neighboring neurons during the adaptation process. The term $D_{b,k}$ refers to the distance between the positions of neurons $b$ and $k$ on the map, while $\delta(t)$ is the neighborhood radius, which typically decreases over time. 

After each epoch, the learning rate and the neighborhood radius are updated as training progresses to ensure convergence to a solution and an accurate representation of the input data, preserving its topology. A common approach to update these parameters is through an exponential decay rate, as described in Equations \ref{eq_5} and \ref{eq_6}:

\begin{equation}
\alpha(t) = \alpha_{0} \exp\left(-\frac{t}{T}\right)
\label{eq_5}
\end{equation}

\begin{equation}
\delta(t) = \delta_{0} \exp\left(-\frac{t}{T}\right)
\label{eq_6}
\end{equation}

These decreases in the learning rate and neighborhood radius enable subtle adjustments, progressively enhancing the Self-Organizing Map's capacity to capture complex topological structures from astronomical data.

\section{Semi-Supervised Learning}

The self-training algorithm involves the iterative application of supervised models to predict new data points \citep{Scudder_1965, Agarwal_1970}. At each training step, data points are labeled according to a selection function, and the supervised model is subsequently retrained using this newly labeled data to refine its predictions \citep{Devgan_2020}.

Semi-supervised learning (SSL) leverages a small proportion of well-labeled data (i.e., data with spectroscopic redshift), to enhance model performance when estimating new labels, which we refer to as pseudo-labeled data \citep{Devgan_2020}. The pseudo-labeled data is gradually and carefully increased during training to minimize learning errors. This iterative selection method enables the model to infer the underlying topology of the data \citep{Alpaydin_2004}.

\subsection{Classification of semi-supervised methods}

The work of \cite{Jesper_2019} proposes a two-level classification for semi-supervised methods. At the first level, methods are categorized into inductive and transductive approaches. Inductive methods focus on constructing a classification model by integrating labeled and unlabeled data. In contrast, transductive methods are primarily concerned with labeling (assigning a spectroscopic redshift) the unlabeled data through predictions. The second level applies exclusively to inductive methods and distinguishes how these methods incorporate unlabeled data into the training process, leading to three distinct types of inductive methods: unsupervised processing, wrapper methods, and intrinsically semi-supervised.

\subsection{Co-training}

Co-training is an inductive semi-supervised learning paradigm \citep{Blum_1998}, where multiple classifiers are trained simultaneously, each using distinct independent subsets of the available training data \citep{Chen_2020}. The fundamental premise of co-training is that the subsets of data contain complementary features (magnitudes and colors), enhancing the learning process of each classifier.

Initially, the dataset with spectroscopic redshift is partitioned into two subsets, each used to train one classifier independently. In subsequent training iterations, the classifiers exchange pseudo-labeled instances based on a defined selection strategy. A key challenge in co-training is confirmation bias \citep{Xu_2023, Arazo_2020}, which refers to the tendency of classifiers to reinforce their erroneous assumptions, leading to the generation of inaccurate or noisy labels. When these biased pseudo-labels are subsequently used for retraining, the accumulated errors can degrade the overall classification or estimation performance.

\begin{figure}[h!]
    \centering
    \includegraphics[width=0.45\textwidth]{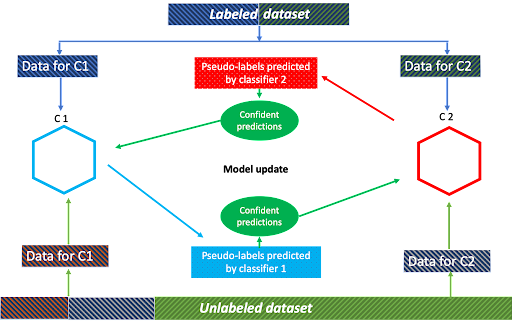}
    \caption{An overview of the co-training process is as follows: both classifiers (C1 and C2) are trained using half of the labeled data set each. Subsequently, in the following iterations, both models are updated with pseudo-labeled data generated through the selection model from the unlabeled instances.}
    \label{co_training} 
\end{figure}

The approaches that mitigate noisy labels exchange data according to a specific selection criterion, complementing each other to enhance the robustness of the training process and improve accuracy by effectively combining labeled and unlabeled data \citep{Du_2020, Hu_2020, Wang_2021}. Empirical studies have shown that co-training requires relatively few iterations to yield high-performing classifiers \citep{Hunter_2022}, as collaborative interaction between models facilitates the rapid development of accurate classifiers. A graphical workflow of the co-training process is presented in Figure \ref{co_training}.

\section{Data}

To test the methodology, we performed a query on the Sloan Digital Sky Survey (SDSS) Data Release 18 (DR18) sample \citep{DataR_2018}. The details of the selection procedure using the CasJobs interface are provided in Appendix A. The sample was restricted to galaxies with redshift values in the range $0.005 < z < 1.0$, ensuring the inclusion of objects with complete photometric measurements across all five SDSS bands ($u, g, r, i, z$) and their corresponding colors ($u-g, g-r, r-i, i-z$).

To minimize dust reddening and ensure reliable morphological measurements, the sample was restricted to low-inclination galaxies ($deV AB_{r}>0.4$), corresponding to inclination angles $>66^{\circ}$, thereby avoiding highly inclined, dust-obscured galaxies. In addition, unresolved sources were removed using the criterion $(petrosianRadius/psfFWHM_r) > 1$, including only objects with complete photometry.

With the selection criteria, 2,155,735 galaxies were selected from the SDSS database. However, a two-stage process was used to refine the training set and eliminate potential outliers.

\subsection{First stage of pre-processing}

\begin{figure}[!ht]
\centering
\includegraphics[scale=0.1]{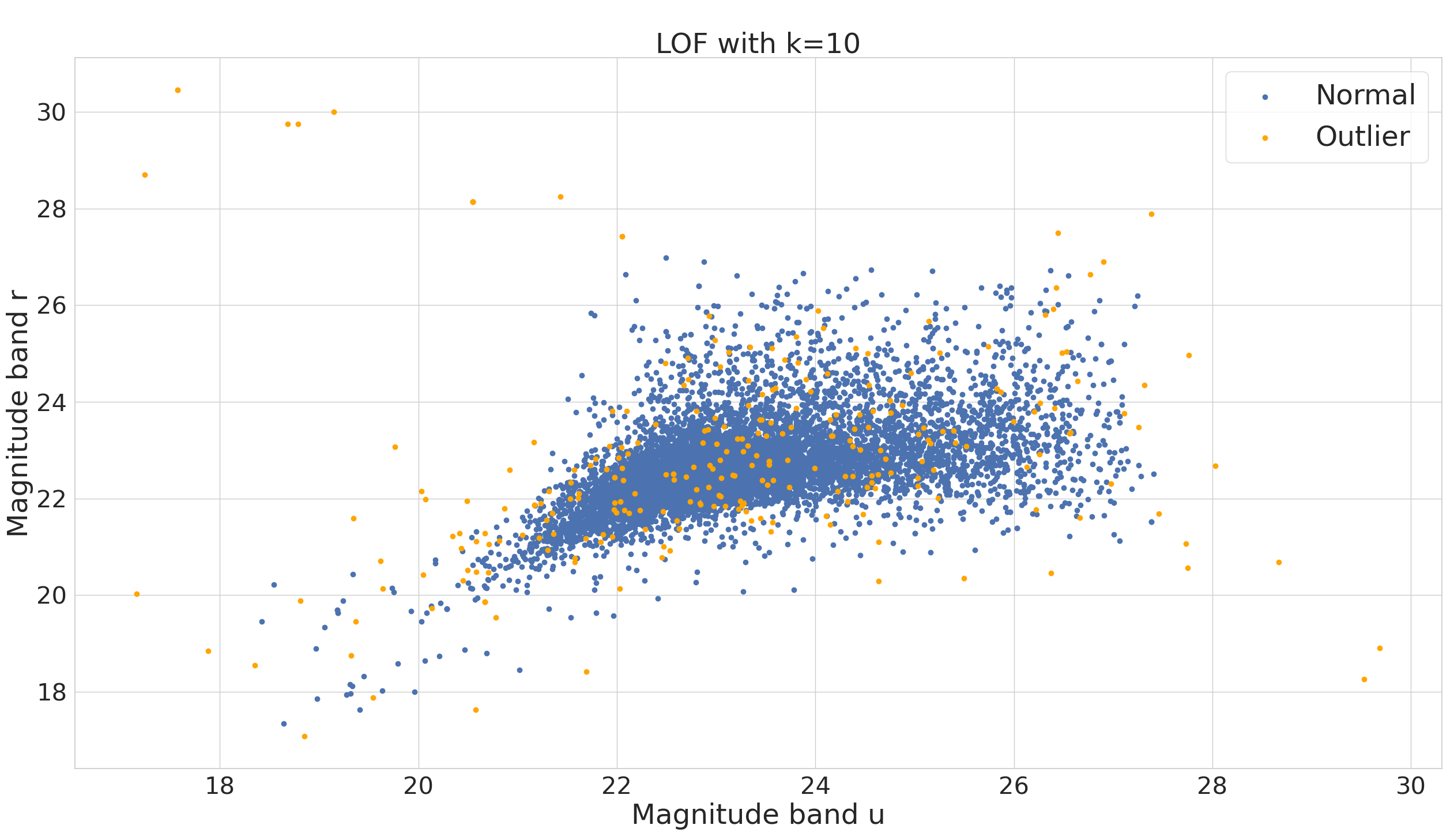}
\caption{Outliers identified using the Local Outlier Factor (LOF) algorithm, with k = 10 neighbors. }
\label{filtro_lof}
\end{figure}

In the first stage, 25,345 galaxies were removed using the Local Outlier Factor (LOF) algorithm \citep{Breuning_2000}. This was achieved by comparing the local density K of each galaxy in the observable feature space with that of its neighbors. Different neighborhood sizes were tested ($K=5, 10, 20, 50$), with $K=10$ providing the best performance in filtering data points located at the boundaries of the densest clusters, as shown in Figure \ref{filtro_lof}.

\subsection{Second stage of pre-processing}

In the second stage, 54,000 galaxies were removed using a criterion based on the standard deviation ($>3\sigma$) to remove galaxies beyond the mean range, thus defining the cluster's boundary. Following this filtering process, 2,076,390 galaxies remained for the subsequent training and validation phases.

\section{Co-SOM}

\begin{figure}[h!]
    \centering
    \includegraphics[width=0.49\textwidth]{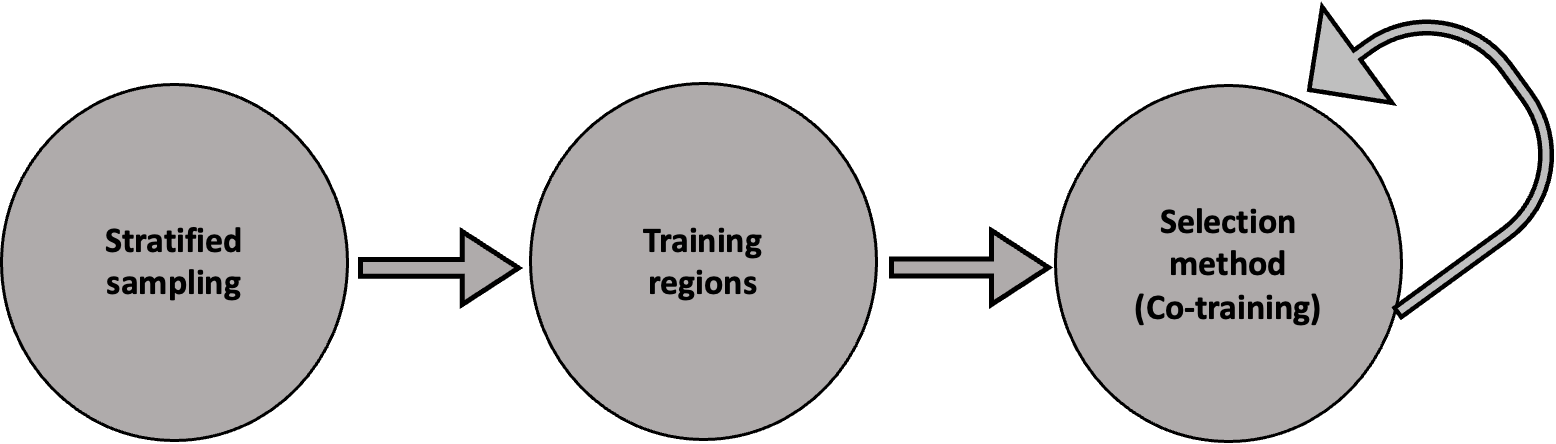}
    \caption{The methodology workflow comprises three primary stages: stratified sampling, training regions, and the selection method of co-training. These stages are carefully designed to address the essential components of the methodology.}
    \label{stages}
\end{figure}

In this work, a methodology based on the co-training of Self-Organizing Maps (SOMs) was implemented to estimate the redshift of galaxies. As previously mentioned, extensive surveys like LSST will provide multi-band photometry of millions of galaxies, each of which must have an associated redshift with high precision ($\Delta z<0.003$) to ensure that the sample's cosmic noise does not dominate the estimation. The redshift estimation can have different sources of uncertainty that contribute to the result, such as random initialization of the models, selection of a training model, lack of data in the training sample, data incompleteness, among others \citep{Sadeh_2016}. Consequently, a multi-stage methodology was developed to minimize the impact of these sources of uncertainty, as illustrated in Figure \ref{stages}.

\subsection{Stratified sampling}
\begin{figure}[h!]
    \centering
    \includegraphics[width=0.49\textwidth]{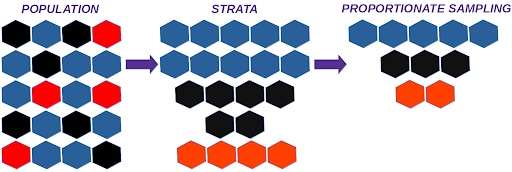}
    \caption{Stratified sampling creates sub-samples while preserving the proportion of the original population.}
    \label{estratificado}
\end{figure}

In the first stage, a stratified sampling technique \citep{Kish_1965, Cochran_1977} was employed to construct training and validation samples from the data. Without proper sampling, especially in the case of imbalanced distributions, higher error rates can arise in regions with fewer data points (galaxies). Figure \ref{estratificado} shows how this technique partitions data into homogeneous subgroups, known as strata, and subsequently draws a random sample from each stratum, ensuring that each stratum is proportionally represented in the final sample.

\subsection{Training regions}
\begin{figure}[h!]
    \centering
    \includegraphics[width=0.49\textwidth]{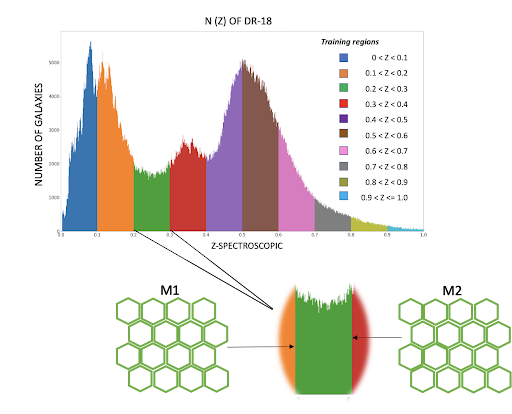}
    \caption{Division of the DR-18 SDSS sample \cite{DataR_2018} into training regions, each represented by a different color. Two Self-Organizing Maps (SOMs) are trained for each region following the co-training method.}
    \label{etapa_1}
\end{figure}

Although stratified sampling preserves the distribution shape across training and validation sets, it does not mitigate the effects of data scarcity at high redshifts. To address this, the sample was divided into $Z$ regions (i.e., by applying cuts in redshift), and separate training maps were generated for each region regardless of the number of data points within a given cut. Figure \ref{etapa_1} illustrates the partitioning of the DR-18 sample, where two maps per region were employed following the co-training methodology.

\subsection{Selection method}

The core functionality of the co-training method lies in its selection strategy, which determines the next set of training data for each subsequent iteration, thereby improving the model's performance. However, one of the main challenges facing semi-supervised techniques is the presence of noisy labels, which can significantly degrade the method's accuracy. To address this, a novel selection approach was developed that considers the topology of the Self-Organizing Maps (SOMs), the inherent features of the data, and the spectroscopic values derived from previous iterations. This methodology aims to minimize the impact of noisy labels during the selection phase. As detailed in Equation \ref{eq_7}, these three factors are integrated into a single metric that evaluates and weights the reliability of labeled and unlabeled data.

\begin{equation}
D(x_{i}) = D_{unlabel}(x_{i}, N) + \beta D_{label}(N)
\label{eq_7}
\end{equation}

In this equation, $x_{i}$ represents an unlabeled data point (e.g., a galaxy without a spectroscopic redshift), and N denotes the neighborhood of the Best Matching Unit neuron (comprising six neighboring neurons). The term $D_{unlabel}$ quantifies the similarity in the feature space, while $D_{label}$ evaluates the similarity among the labels of neighboring neurons. In addition, the parameter $\beta$ is introduced to control the influence of $D_{label}$ on the overall metric, as indicated in Equation \ref{eq_8}. The distance $D_{unlabel}$ is defined as follows:

\begin{equation}
D_{unlabel}(x_{i}, N) = \sum_{j=1}^{6} d(x_{i}, N_{j})
\label{eq_8}
\end{equation}

In this equation, $D_{unlabel}(x_{i}, N)$ computes the sum of the distances between the unlabeled data point $x_{i}$ and the six neurons in the neighborhood $N$. It is important to note that a lower value of $D_{unlabel}$ indicates a greater similarity between the input data and the Best Matching Unit. Next, we define $D_{label}$ as follows in Equation \ref{eq_9}:

\begin{equation}
D_{label}(N) = \sum_{j=1}^{6} d(\bar{y}, y_{j})
\label{eq_9}
\end{equation}

In this equation, $D_{label}(N)$ calculates the sum of the distances between the average label $\bar{y}$ of the neighboring neurons and the label $y_{j}$ of the BMU. A lower value of $D_{label}$ signifies a higher similarity between the input data and the BMU, based on the labels of the neighboring neurons. To evaluate the overall consistency between the unlabeled data and the trained models, both $D_{unlabel}$ and $D_{label}$ are considered in the selection method, as shown in Equation \ref{eq_7}. 

\begin{figure}[h!]
  \centering
   \includegraphics[scale=0.20]{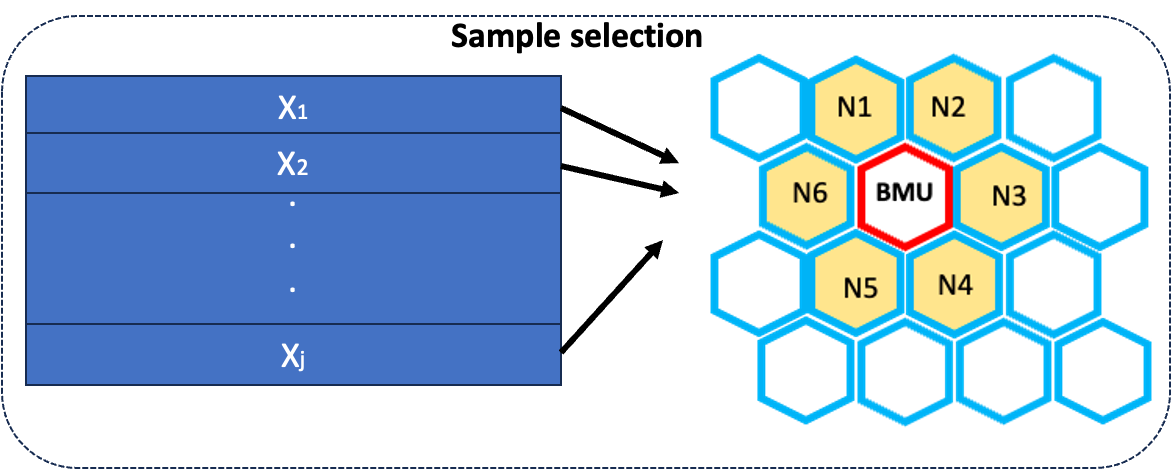}
  \caption{To evaluate the similarity between labeled and unlabeled data at each iteration of Co-SOM, the topological properties of the hexagonal map architecture are exploited.}         
  \label{selection}                            
\end{figure}

Figure \ref{selection} illustrates the general concept of evaluating this consistency by comparing the similarity between the unlabeled data $X_{j}$ and the trained model, incorporating both the feature space and the label information.

By incorporating both feature space and label consistency, this approach minimizes the impact of noisy labels. It ensures that the pseudo-labeled data added to the model are consistent with the underlying structure of the map. This results in a more robust and accurate model that can better predict the redshift of galaxies with minimal supervision.

When the value of $D(x_{i})$ decreases, the pseudo-label becomes more accurate and consistent. However, evaluating pseudo-labeled data only within the current iteration does not guarantee stability or convergence, as the process may still be susceptible to noisy labels, which can potentially introduce confirmation bias into the predictions. To address this issue we adopt the aggregation method proposed by \citet{Xu_2023}, defined in Equation \ref{eq_10}.

This approach stabilizes the pseudo-labeling process by weighting the previous and current prediction, thereby enforcing consistency in the selection mechanism. In practice, if a high-quality prediction remains stable across iterations, it is more likely to be selected, which mitigates the effects of confirmation bias.

\begin{equation}
\Theta^{(t)}(x_{i}) = (1-\gamma)\Theta^{(t-1)}(x_{i}) + (\gamma D^{(t)}(x_{i}))
\label{eq_10}
\end{equation}

Equation \ref{eq_10} describes the aggregation method that updates the model parameters based on the previous predictions and the current model output, where:

\begin{itemize}
    \item $\Theta^{(t)}(x_{i})$ represents the updated pseudo-label for data point $x_{i}$ at iteration t
    \item $\Theta^{(t-1)}(x_{i})$ is the pseudo-label assigned to $x_{i}$ at the previous iteration $t-1$
    \item $D^{(t)}(x_{i})$ is the current prediction or output (based on the current iteration’s model)
    \item $\gamma$ is a hyperparameter that controls the balance between the last prediction and the new one. Its value varies between 0 and 1, determining the extent to which the previous iteration's prediction should influence the current iteration.
\end{itemize}
 
When $\gamma$ is set to a value closer to 1, the most recent prediction has more weight, whereas a value closer to 0 places more weight on the previous label. The introduction of this parameter enables flexibility in controlling the smoothing of pseudo-label updates, providing a mechanism to manage model stability.

The behavior of $\Theta^{(t)}(x_{i})$ over iterations is crucial to ensuring the consistency of the model. A decreasing value of $\Theta^{(t)}(x_{i})$ indicates that the model output stabilizes over iterations, signifying that the pseudo-labels are increasingly reliable. In this way, the aggregation process contributes to the stability of the co-training process, improving the accuracy of the predictions over time.

To integrate the aggregation method and the metric that quantifies similarity in the feature space and evaluates similarity between neighboring neurons' labels, as presented in Equation \ref{eq_7}, a probability metric was defined to select new unlabeled data, as shown in Equation \ref{eq_11}.

\begin{equation}
 p(x_{i}) =\frac{max_{x_{i}} \Theta^{(t)}(x_{i})  - \Theta^{(t)}(x_{i})}{\sum_{x \in x_{u}} max_{x_{i}} \Theta^{(t)}(x_{i})  - \Theta^{(t)}(x_{u})}
\label{eq_11}
\end{equation}

Although a probability has been defined to select new training data, the concept of a Minimum Quality Criterion (MQC) was introduced, as detailed in Equation \ref{eq_12}. This criterion regulates the number of data points selected in each iteration, ensuring that only those data points that meet the minimum similarity threshold are chosen and labeled for training in the subsequent iteration of the method.

\begin{equation}
MQC \geq  max_{x_{i}} p(x_{i}) * Threshold
\label{eq_12}
\end{equation}

This criterion further helps to reduce the impact of noisy labels introduced by incorrect pseudo-labels during the selection process. The parameter $threshold$ controls the similarity between the data; the higher the parameter, the more restrictive and selective the selection method becomes. This ensures more reliable and accurate pseudo-labels, which are crucial when working with large, high-dimensional datasets, such as those used in redshift prediction.

\subsection{Performance metrics}
To evaluate the performance of the methodology, we use six metrics. The first three are used in regression problems: Mean squared error (MSE), mean absolute error (MAE), and R-squared score ($R^{2}$). where $n$ represents the total number of samples, $\hat{y_{i}}$ denotes the estimated value for the i-th sample, $y_{i}$ is the true value and $\bar{y}$ is the mean of the true values. These metrics are calculated as follows \ref{eq_13}, \ref{eq_14}, \ref{eq_15}:

\begin{equation}
MSE =\frac{1}{n} \sum_{i=1}^{n} (y_{i}-\hat{y_{i}})^{2}
\label{eq_13}
\end{equation}

\begin{equation}
MAE =\frac{1}{n} \sum_{i=1}^{n} |y_{i}-\hat{y_{i}}|
\label{eq_14}
\end{equation}

\begin{equation}
R^{2} = 1 - \frac{\sum_{i=1}^{n} (y_{i}-\hat{y_{i}})^{2}}{\sum_{i=1}^{n} (y_{i}-\bar{y})^{2}}
\label{eq_15}
\end{equation}

In addition, three other metrics are used to evaluate the performance of photometric estimations: bias, precision, and fraction of outliers based on the work of \cite{Henghes_2021} and \cite{Ilbert_2006}. These metrics are calculated as in equations: \ref{eq_16}, \ref{eq_17}, \ref{eq_18}:

\begin{equation}
bias (\Delta z) =  \frac{z_{phot}-z_{spec}} {1+z_{spec}}
\label{eq_16}
\end{equation}

\begin{equation}
\sigma_{zp} = 1.48*median(|\Delta z|)
\label{eq_17}
\end{equation}

\begin{equation}
out\_frac = \frac{N(\Delta z)>0.1}{N_{Total}}
\label{eq_18}
\end{equation}

where $N$ represents the total number of samples, $z_{phot}$ is the photometric estimations and $z_{spec}$ denotes the spectroscopic values.

\subsection{Experiments}

Preliminary tests were performed to identify the optimal SOM configuration for this dataset in Table \ref{tab:SOM}. We found that using a large initial neighborhood radius ($\approx50\%$ of the map size) combined with a high initial learning rate ($\alpha = 1$) improves convergence.

In subsequent iterations, the neighborhood radius was gradually reduced to $10\%$ of the map size in the second iteration and to $10\%$ of its previous value thereafter, allowing the model to capture subtle changes introduced by unlabeled data. This strategy improved redshift estimation accuracy and reduced the required co-training to an average of 10 rounds, while maintaining stability across all six performance metrics (both regression accuracy and the quality of redshift estimation). The number of epochs was fixed at 200, as no significant improvements were observed beyond this value; this parameter was therefore adopted as a stopping criterion together with the neighborhood radius size ($\delta(t) > 1$). It is important to note that the SOM weights were continuously updated between iterations to refine the patterns learned throughout the rounds.

\begin{table}[h!]
\centering
\caption{Evaluation of SOM parameters}
\begin{tabular}{ |p{3cm}||p{3cm}| }
 \hline
 \multicolumn{2}{|c|}{Training parameters} \\
 \hline
 $\alpha(t)$ (Learning rate)  & 0.1 -- 0.5 -- 1.0  \\
  \hline
  $\delta (t)$ (Radius)  & 10\% -- 50\% -- 75\%  \\
  \hline
  Epochs & 100 -- 200 -- 500 \\
  \hline
 \end{tabular}
 \label{tab:SOM}
\end{table}

\begin{figure}[h!]
    \centering
    \includegraphics[width=0.5\textwidth]{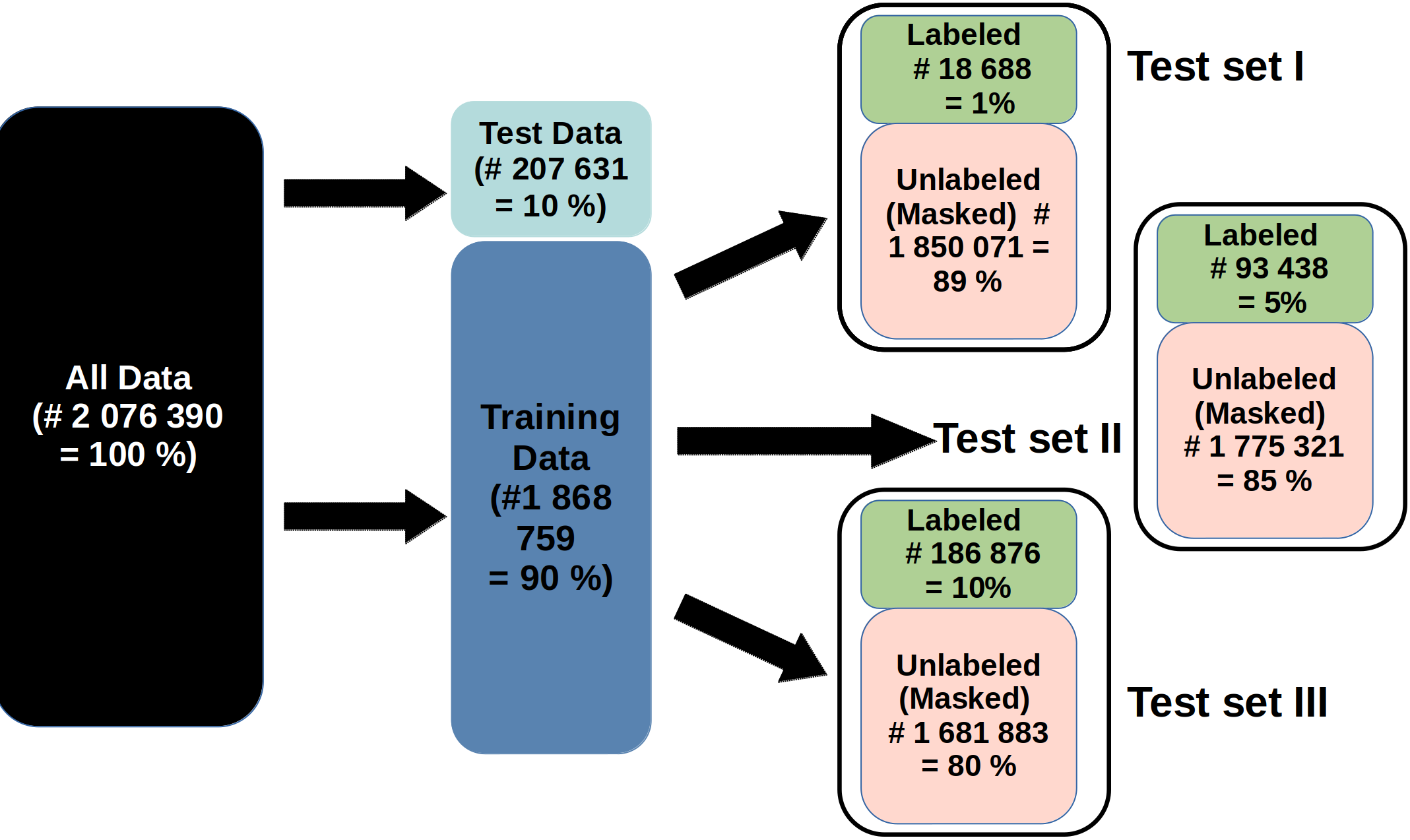}
    \caption{Evaluation of the Co-SOM method performance across three test sets.}
    \label{tests_1}
\end{figure}

To evaluate the Co-SOM method, three independent test sets were created based on the amount of data used during the training phase: 1\%, 5\%, and 10\%. Each test set was analyzed from multiple perspectives, making it essential to select optimal parameters for each case. All experiments were conducted on a 64-bit server equipped with an Intel Core i7-8700 CPU (6 cores, 12 threads, 3.2–4.6 GHz) and 64 GB of RAM. Figure \ref{tests_1} shows the percentages of labeled and unlabeled data for each test set, while the outputs derived from the maps correspond to the actual redshift estimates obtained by the method.

To avoid bias in selecting the best model from each test set, a K-fold cross-validation was performed (K = 5), a process similar to that employed in supervised methods. For each set, different parameters associated with co-training, Self-Organizing Maps, and training regions were evaluated. These parameters are summarized in Table \ref{tab:hyper}.

\begin{table}[h!]
\centering
 \caption{Evaluation of parameters using SDSS Data Release 18}
\begin{tabular}{ |p{3cm}||p{3cm}| }
 \hline
 \multicolumn{2}{|c|}{Training parameters} \\
 \hline
 $\beta$ (Co-training)  & 0.1 -- 0.5 -- 1.0  \\
  \hline
  $\gamma$ (Co-training)  & 0.1 -- 0.5 -- 1.0  \\
  \hline
  [MS] MAP SIZE (SOM)& 20x25 -- 30x35 \\
  \hline
   \# Z CUTS (Training regions)& 2 -- 5 -- 10 \\
  \hline
 \end{tabular}
\label{tab:hyper}
\end{table}

Although the methodology was tested based on the percentage of labeled data used during training, it is also essential to analyze the effect of incorporating more data during co-training through the selection method. To this end, experiments were carried out by varying the co-training selection threshold, as shown in Equation \ref{eq_12}. This threshold was initially set to 95\% for the three previous test sets to prevent bias and instability caused by noisy labels. However, the effects of lowering the threshold to 75\%, 55\%, and 45\% were also analyzed to determine if there is a lower limit on the quality of the labels to avoid training with noisy labels.

\section{Results and discussion}

The methodology was evaluated using three parameters to provide a comprehensive analysis from different perspectives: model training (co-training), the effect of map size (SOM), and the impact of data scarcity at higher redshifts (training regions). Three sets of tests were performed by varying the proportion of labeled spectroscopic data used during training at 1\%, 5\%, and 10\%, thereby testing the operational limits of the method. This approach is particularly relevant, as upcoming large-scale surveys will similarly involve large datasets with a limited amount of labeled (spectroscopic) data.

Table \ref{tab:results_1} presents the results obtained from Test Set I, where the co-training selection method was evaluated under different values of the parameters $\beta$ and $\gamma$; the best performance for each metric is highlighted in red. It is worth noting that in this work $\beta$ was fixed globally (tested in the range 0.1–1.0). The results show that performance improves when $\beta$ assumes its maximum value, as this parameter weights the relevance of labels (spectroscopy) and observables (magnitudes and colors) in equation \ref{eq_7}.

Furthermore, a value of $\gamma$ between 0.5 and 1.0 improves the precision of the estimation. This parameter adjusts the probability of labeling a data point based on its reliability, as assigned in a previous training phase. Specifically, suppose a data point was initially a strong candidate for labeling but failed to meet the selection criterion during the current training session. In that case, it is more likely to satisfy the criterion and be labeled in the following iteration.

\begin{table*}[ht!]
\centering
\caption{Results of the experimental evaluation}
\resizebox{\textwidth}{!}{%
    \begin{tabular}{|c|c|c|c|c|c|c|c|c|}
    \hline
    \multicolumn{9}{ |c| }{\textbf{Test Set I with 1\% of labeled data (spectroscopic data)}}\\
    \hline
    \textbf{Metrics}&\textbf{T01}&\textbf{T02}&\textbf{T03}&\textbf{T04}&\textbf{T05}&\textbf{T06}&\textbf{T07}&\textbf{T08}\\
    \hline
  $\beta$ (eq. \ref{eq_7}) &$\beta = 0.1$ &$\beta = 0.5$ & $\beta = 1.0$ & $\beta = 1.0$ & $\beta = 1.0$ & $\beta = 1.0$ & $\beta = 1.0$& $\beta = 1.0$ \\
  $\gamma$ (eq. \ref{eq_10})&$\gamma = 0.1$  & $\gamma = 0.1$& $\gamma = 0.1$ &$\gamma = 0.5$&$\gamma = 1.0$ & $\gamma = 1.0$ & $\gamma = 1.0$& $\gamma = 1.0$  \\
  MS (SOM Map Size)&MS = 20x25  & MS = 20x25 & MS = 20x25  &MS = 20x25 & MS = 20x25 &MS = 30x35 &MS = 30x35& MS = 30x35  \\

  \# CUTS (Z training regions)& \# CUTS = 2  &  \# CUTS = 2 &  \# CUTS = 2  & \# CUTS = 2 &  \# CUTS = 2 & \# CUTS = 2 & \# CUTS = 3 & \# CUTS = 2\\
 
\hline 
\textbf{MAE}&0.04698 $\pm$ 0.00041& 0.04712 $\pm$ 0.00076& 0.04700 $\pm$ 0.00047& 0.04659 $\pm$ 0.00023& 0.04675 $\pm$ 0.00023&\textcolor{red}{0.04449 $\pm$ 0.00022}& 0.04689 $\pm$ 0.00051& 0.04720 $\pm$ 0.00026\\
\hline
\boldmath{$R^{2}$} & 0.87285 $\pm$ 0.00530& 0.87265 $\pm$ 0.00978& 0.87234 $\pm$ 0.00608& 0.87762 $\pm$ 0.00532& 0.87831 $\pm$ 0.00465&\textcolor{red}{0.88501 $\pm$ 0.00375}& 0.87475 $\pm$ 0.00708& 0.86078 $\pm$ 0.00197\\ 
\hline

\textbf{MSE}& 0.00585 $\pm$ 0.00024& 0.00586 $\pm$ 0.00045& 0.00587 $\pm$ 0.00028& 0.00563 $\pm$ 0.00024& 0.00560 $\pm$ 0.00021&\textcolor{red}{0.00529 $\pm$ 0.00017}& 0.00576 $\pm$ 0.00033& 0.00624 $\pm$ 0.00009\\
\hline

\boldmath{$\Delta z$}& 0.00072 $\pm$ 0.00073& 0.00048 $\pm$ 0.00065& 0.00052 $\pm$ 0.00054& 0.00031 $\pm$ 0.00071&\textcolor{red}{0.00007 $\pm$ 0.00022}& 0.00103 $\pm$ 0.00020& 0.00039 $\pm$ 0.00056& 0.00352 $\pm$ 0.000380
\\
\hline

\boldmath{$\sigma_{zp}$}&0.03475
 $\pm$ 0.00031& 0.03485 $\pm$ 0.00018& 0.03474 $\pm$ 0.00029& 0.03472 $\pm$ 0.00022& 0.03485 $\pm$ 0.00034& \textcolor{red}{0.03241 $\pm$ 0.00017}& 0.03477 $\pm$ 0.00027& 0.03333
 $\pm$ 0.00012
 \\
\hline
\textbf{Out\_Frac}&0.02221 $\pm$ 0.00083& 0.02265 $\pm$ 0.00177& 0.02232 $\pm$ 0.00093& 0.02169 $\pm$ 0.00056& 0.02144 $\pm$ 0.00103& \textcolor{red}{0.02083 $\pm$ 0.00027}& 0.02206 $\pm$ 0.00072& 0.02675 $\pm$ 0.00085
\\
\hline
\end{tabular}}
\label{tab:results_1}
\end{table*}

Similarly, parameters related to the map size and the number of training regions were evaluated. The results indicate that doubling the number of neurons (map size) improved almost all metrics. This is attributed to the fact that, in a larger map, galaxies are distributed more evenly across a greater number of neurons during training. However, the bias ($\Delta z$) is negatively impacted because larger maps decrease the number of galaxies for the neuron, and the uncertainty in redshift estimation increases.

Alternatively, increasing the number of training regions resulted in a decrease in accuracy, as the combination of limited data and the over-representation of areas with sparse data affects the resolution of the estimation. However, it is crucial to highlight that the method achieves highly accurate results even with only $1\%$ of the data results ($\Delta z = 0.00033$), which is well below the bias threshold defined in the LSST-DESC Science Requirements Document (SRD) \citep{LSST_2021}. This demonstrates that systematic errors in the estimation are not dominated by background noise from the LSST cosmological sample.

\begin{figure}[h!]
    \centering
    \includegraphics[width=0.40\textwidth]{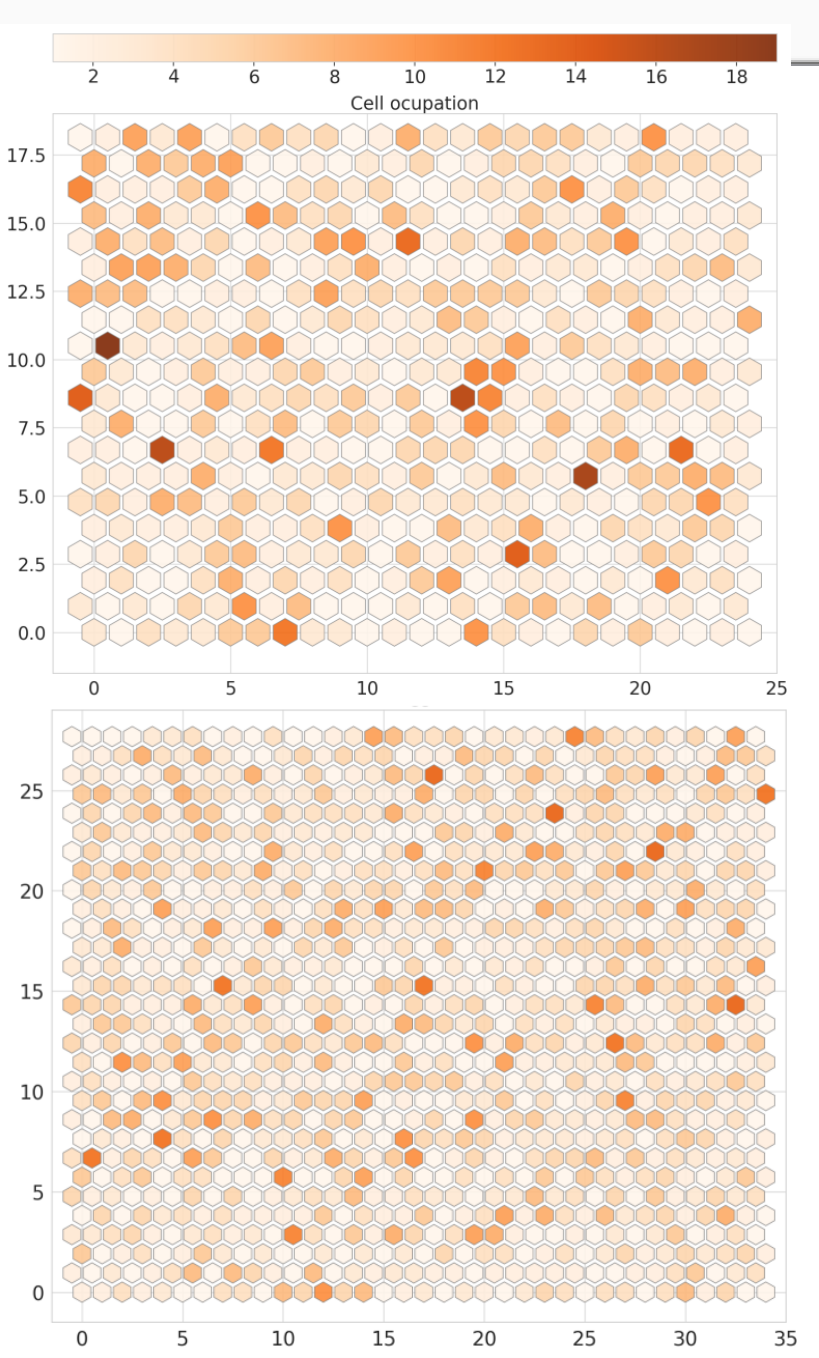}
    \caption{Both maps illustrate the cell occupancy of the Self-Organizing Maps (SOMs). The upper map corresponds to Test T05, which contains 500 cells, whereas the lower map, derived from Test T06, consists of 1,050 cells.}
    \label{ocupation}
\end{figure}

Figure \ref{ocupation} illustrates the distribution of the data used for training after co-training, quantifying the number of galaxies per cell (neuron) in tests T05 and T06. The lower map corresponds to test T06, which has double the number of cells, allowing for a more uniform distribution of galaxies within the map, with an average of 3 to 6 galaxies per cell. However, neurons with no assigned redshift value and with fewer galaxies per cell are observed, so the average redshift assignment is more influenced by outliers, introducing greater uncertainty, as reflected in Table \ref{tab:results_1}. Consequently, map size is a critical parameter for model precision, as larger maps result in increased uncertainty and higher computational cost during training.

In addition to the qualitative assessment, quantitative measurements of training times were performed as a function of map size and the amount of labeled data. Table \ref{tab:comp_time} summarizes these results, reporting the average runtime per co-training round and the total runtime for complete training runs (10 rounds). As an illustration, with a 30x35 map and $10\%$ labeled data ($200,000$ objects), a full run required approximately 101 minutes, whereas with a 40x40 map, the runtime increased to nearly 120 minutes. These values reflect the already parallelized CPU implementation. Nevertheless, scaling to datasets at the level of LSST (billions of galaxies) would demand further improvements, such as distributing the workload across multiple servers or exploiting GPU acceleration. Such strategies could reduce runtime and make it feasible at the survey scale.

\begin{table}[ht!]
\centering
\caption{Average runtimes per round and per full run for two SOM map sizes as a function of the fraction of spectroscopic data used}
\label{tab:comp_time}
\resizebox{\columnwidth}{!}{%
\begin{tabular}{|c|c|c|c|}
\hline
\textbf{Map Size} & \textbf{Spectroscopic fraction (objects)} & \textbf{Minutes / round} & \textbf{Minutes / full run} \\
\hline
$30 \times 35$ & 1\% (20,000)  & 2.13  & 21.30  \\
$30 \times 35$ & 5\% (100,000) & 5.47  & 54.76  \\
$30 \times 35$ & 10\% (200,000) & 10.11 & 101.26 \\
\hline
$40 \times 40$ & 1\% (20,000)  & 2.81  & 28.18  \\
$40 \times 40$ & 5\% (100,000) & 7.91  & 79.23  \\
$40 \times 40$ & 10\% (200,000) & 11.97 & 119.94 \\
\hline
\end{tabular}%
}
\end{table}

\begin{figure}[!ht]
    \centering
    \includegraphics[width=0.48\textwidth]{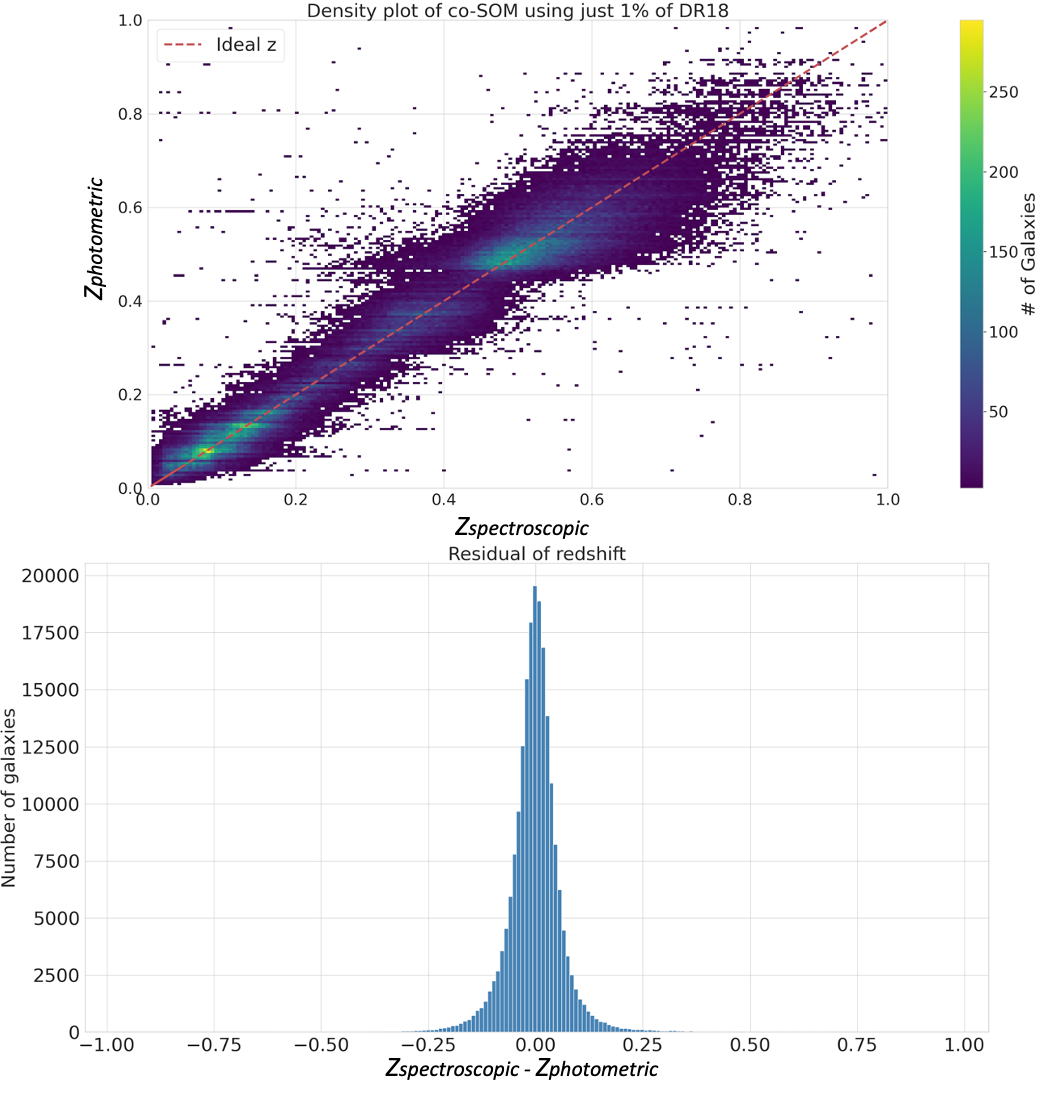}
    \caption{ The upper panel shows the density predicted ($Z_{predicted}$) vs spectroscopic redshift ($Z_{spectroscopic}$) in T06. The lower panel displays the residual vector of $\Delta z$ in a histogram.}
    \label{densityandresidual}
\end{figure}

Figure \ref{densityandresidual} presents two graphs showing the results of Test T06. The upper map shows the density map of the estimated redshift versus the spectroscopic redshift, where the red dashed line represents the ideal value of Z ($\Delta z=0$). In the lower graph, the histogram of the residual redshift vector is shown, which is defined as $residual=Z_{spectroscopic} - Z_{photometric}$ \citep{Hoyle_2016}.

In the upper panel, the density peaks around $Z \approx 1.5$ and $Z \approx 7.5$, corresponding to the eBOSS and Main Galaxy Sample surveys \citep{Howlett_2015,Talbot_2021}, which contribute most of the data. For $Z > 0.8$, the data are sparser, resulting in greater scatter. Residuals were characterized by computing the median and $\sigma_{68}$ of $\Delta z$, yielding $\mu=0.001362$ and $\sigma_{68}=0.07133$, indicating that the model's average bias is negligible. The residual distribution is strongly concentrated around $\mu \approx 0$ (lower panel), with most of the sample exhibiting a maximum scatter of $\pm 0.25Z$ and no discernible systematic trends, indicating that the model is well-calibrated and free from overfitting or underfitting.

However, the tails of the residual distribution include galaxies whose photometric redshifts deviate significantly from their spectroscopic values. These catastrophic, non-symmetric errors can bias cosmic distance estimates and impact large-scale structure studies that assume Gaussian errors. Mitigation strategies, such as robust outlier rejection or explicit modeling of the full residual distribution, should be applied when using these photometric redshifts for precision cosmology.

Two additional test sets, II and III, were conducted to evaluate the model's performance as the proportion of training data increases. In these sets, the parameters $\beta$ and $\gamma$ were fixed, as they exhibited the same behavior observed in the results of set I. As shown in Table \ref{tab:results_2}, performance improves with an increase in map size, as seen in the T10 and T15 tests. This is because a larger amount of data requires the use of larger maps for training. Furthermore, increasing the number of regions leads to a degradation in the precision of the estimation, as indicated by the results of test set I.

\begin{figure}[ht]
    \centering
    \includegraphics[width=0.48\textwidth]{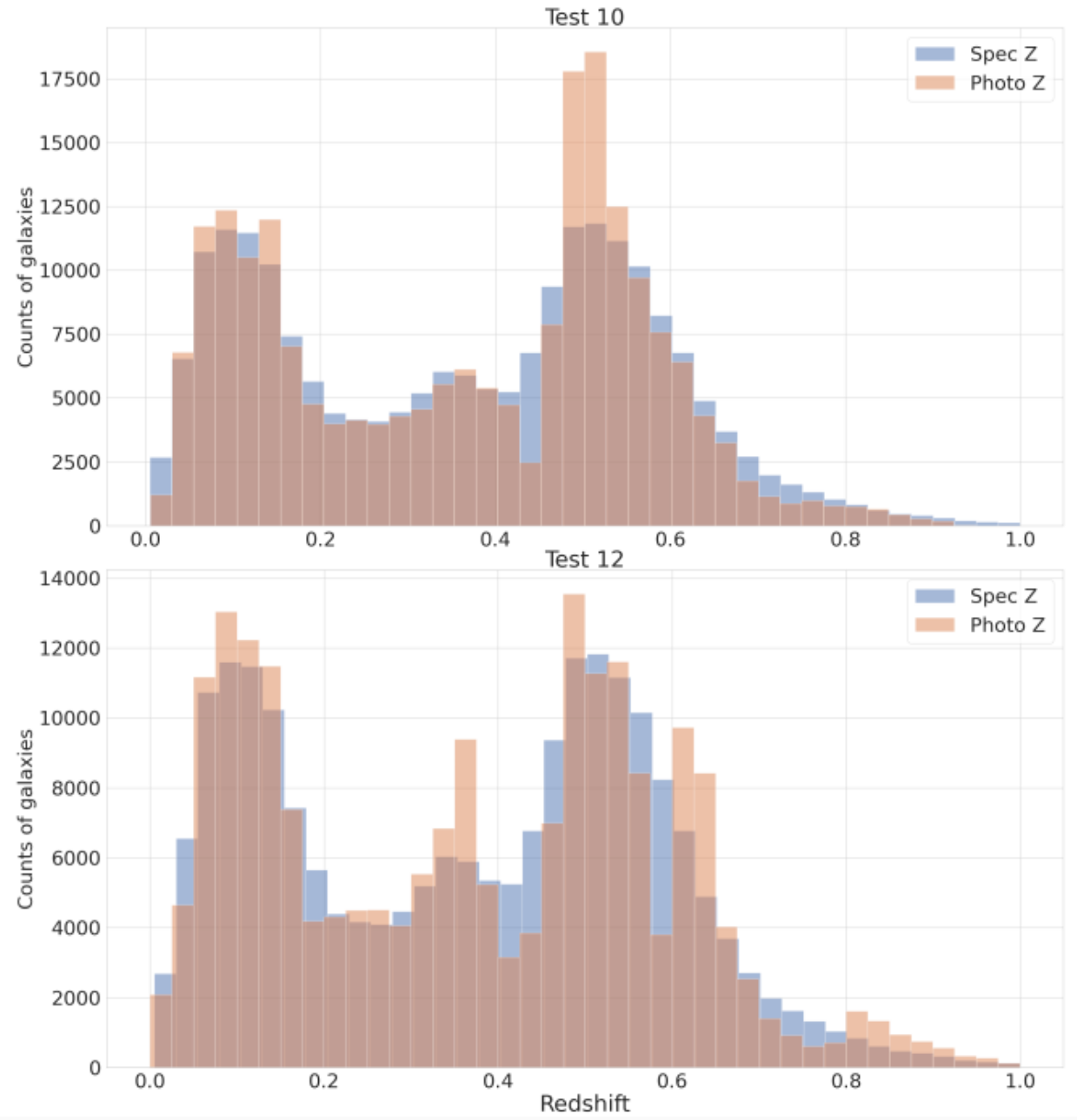}
    \caption{These histograms are from tests 10 and 12. Both histograms show the distribution of $Z_{spectroscopic}$ and estimated in each test.}
    \label{regions}
\end{figure}

Despite the increase in the training regions, the results of the three test sets showed degradation. As illustrated in Figure \ref{regions}, the application of cuts and region subdivisions caused the maps to compensate for areas with a lack of data by over-representing these regions in the final output.

\begin{table*}[!t]
\centering
\caption{Results of the experimental evaluation}
\resizebox{\textwidth}{!}{%
\begin{tabular}{|c|c|c|c|c|c|c|c|}
\hline
\multicolumn{5}{ |c| }{\textbf{Test Set II with 5\% of labeled data}}&\multicolumn{3}{ |c| }{\textbf{Test Set III with 10\% of labeled data}}\\
\hline
\multirow{3}*{\textbf{Metrics}} &T09&T10&T11&T12&T13&T14&T15\\
\hline
  & \textbf{MS = 20x25}  &  \textbf{MS = 30x35 } & MS = 30x25  &MS = 30x35 & \textbf{MS = 20x25}  &  \textbf{MS = 30x35 } & \textbf{MS = 40x40}     \\
  & \# CUTS = 2  &  \# CUTS = 2 &  \textbf{ \# CUTS = 3 } & \textbf{ \# CUTS = 5 }&  \# CUTS = 2 &  \# CUTS = 2  & \# CUTS = 2  \\
 
\hline
\textbf{MAE}&0.04616 $\pm$ 0.00055& \textcolor{red}{0.04328 $\pm$ 0.00028}& 0.04444 $\pm$ 0.00034& 0.04490 $\pm$ 0.00016& 0.04514 $\pm$ 0.00149& 0.04285 $\pm$ 0.00032& \textcolor{red}{0.04145 $\pm$ 0.00017}
\\
\hline
\textbf{$R^{2}$}&0.87954 $\pm$ 0.00783& \textcolor{red}{0.89128 $\pm$ 0.00170}& 0.88395 $\pm$ 0.00340& 0.86910 $\pm$ 0.00190& 0.88421 $\pm$0.01057& 0.89561 $\pm$ 0.00244& \textcolor{red}{0.90207 $\pm$ 0.00147}
\\
\hline
\textbf{MSE}&0.00554$\pm$ 0.00036& \textcolor{red}{0.00500 $\pm$ 0.00008}& 0.00534 $\pm$ 0.00016& 0.00594 $\pm$ 0.00009& 0.00533 $\pm$ 0.00049& 0.00480
 $\pm$ 0.00011& \textcolor{red}{0.00450 $\pm$ 0.00007}
\\
\hline
\textbf{$\Delta z$}&0.00028$\pm$ 0.00102& \textcolor{red}{0.00008 $\pm$ 0.00025}& 0.00346 $\pm$ 0.00040& 0.00444 $\pm$ 0.00043& 0.00022 $\pm$ 0.00064&\textcolor{red}{0.00015 $\pm$ 0.00041}& 0.00021 $\pm$ 0.00020
 \\
\hline
\boldmath{$\sigma_{zp}$}&0.03440 $\pm$ 0.00029& 0.03178 $\pm$ 0.00008& 0.03195 $\pm$ 0.00019& 0.03093 $\pm$0.00020& 0.03414 $\pm$ 0.00018& 0.03169$\pm$ 0.00011& \textcolor{red}{0.03040 $\pm$ 0.00017}
 \\
 \hline
\textbf{Out\_Frac}&0.02095$\pm$ 0.00137& \textcolor{red}{0.01838 $\pm$ 0.00046}& 0.02136 $\pm$ 0.00057& 0.02638 $\pm$ 0.00038& 0.00117 $\pm$ 0.01950& 0.01761
 $\pm$ 0.00090& \textcolor{red}{0.01693 $\pm$ 0.00010}
\\
\hline
\end{tabular}}
\label{tab:results_2}
\end{table*}

\begin{figure}[h!]
    \centering
    \includegraphics[width=0.40\textwidth]{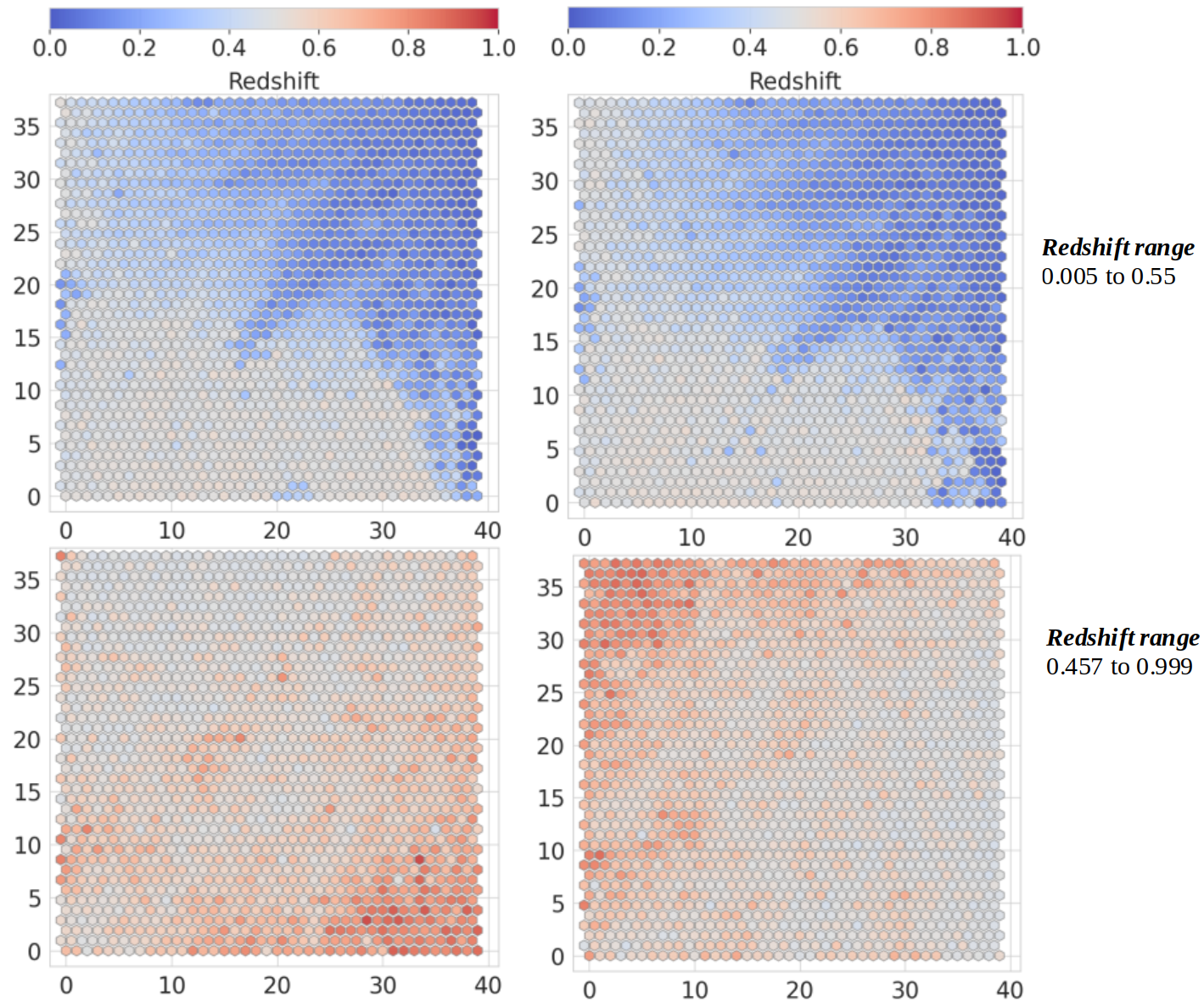}
    \caption{Redshift mapped in four maps for test T15. Most of the neurons map the two peaks of data at $z \approx 0.1$ and $Z \approx 0.6$. The range $Z > 0.8$ has been mapped to a few neurons due to a lack of data. }
    \label{redshift}
\end{figure}

The best results were obtained using 10\% of labeled data, which was achieved in test T15, a finding consistent with observations from previous tests. Since more data were used for training, the model was trained with a map with 1600 neurons to guarantee a uniform distribution of the data. Figure \ref{redshift} illustrates the redshift mapping along the neurons. As shown, most of the neuronal mapping concentrates around the prominent redshift peaks, with a gradual distribution across the maps that separates high from low Z values. In addition, neurons mapped with high Z values are sparse, located primarily at the map boundaries, which is consistent with the amount of input data.

Furthermore, for the best-performing tests from the three test sets (I, II, III), the evolution of pseudo-label accuracy across co-training rounds was evaluated. In all cases, metrics improved continuously as training progressed, as shown in Table \ref{tab:pseudo_label_metrics}. For instance, in T06 (Test Set I), the MAE decreased from 0.070 in round 2 to 0.044 in round 10, $R^{2}$ increased from 0.77 to 0.89, and the outlier fraction dropped from $\approx 6\%$ to $\approx 2\%$.

Similar trends were observed in T10 (Test Set II) and T15 (Test Set III), with MAE decreasing to 0.043 and 0.041, respectively, and outlier fractions falling below $2\%$. The bias $\Delta z$ rapidly converged to values close to zero, indicating that pseudo-labels stabilize quickly. These results show that pseudo-label propagation is robust and effectively improves the quality of unlabeled data over successive training rounds.

\begin{table}[t]
\centering
\caption{Evolution of metrics across co-training rounds for the best performing tests from Test Sets I (T06), II (T10), and III (T15)}
\label{tab:pseudo_label_metrics}
\adjustbox{max width=\columnwidth}{
\begin{tabular}{|c|ccc|ccc|ccc|}
\hline
Round & \multicolumn{3}{c|}{\textbf{MAE}} & \multicolumn{3}{c|}{\boldmath{$R^{2}$}} & \multicolumn{3}{c|}{\textbf{MSE}} \\
\hline
      & T06 & T10 & T15 & T06 & T10 & T15 & T06 & T10 & T15 \\
\hline
2  & 0.0703 & 0.0700 & 0.0624 & 0.7689 & 0.7626 & 0.7421 & 0.01063 & 0.01092 & 0.01186 \\
4  & 0.0684 & 0.0659 & 0.0681 & 0.7975 & 0.8037 & 0.7917 & 0.00932 & 0.00903 & 0.00958 \\
6  & 0.0643 & 0.0647 & 0.0650 & 0.8096 & 0.8115 & 0.8008 & 0.00876 & 0.00867 & 0.00916 \\
8  & 0.0606 & 0.0615 & 0.0415 & 0.8205 & 0.8211 & 0.9010 & 0.00826 & 0.00823 & 0.00455 \\
10 & 0.0445 & 0.0430 & 0.0415 & 0.8850 & 0.8946 & 0.9021 & 0.00529 & 0.00485 & 0.00450 \\
\hline
\end{tabular}
}

\vspace{0.3cm}

\adjustbox{max width=\columnwidth}{
\begin{tabular}{|c|ccc|ccc|ccc|}
\hline
Round & \multicolumn{3}{c|}{\boldmath{$\Delta z$}} & \multicolumn{3}{c|}{\boldmath{$\sigma_{zp}$}} & \multicolumn{3}{c|}{\textbf{Out\_Frac}} \\
\hline
      & T06 & T10 & T15 & T06 & T10 & T15 & T06 & T10 & T15 \\
\hline
2  & -0.0138 & -0.01555 & 0.01681 & 0.05965 & 0.05842 & 0.06020 & 0.0605 & 0.0566 & 0.0569 \\
4  & 0.00578 & 0.00177 & 0.00179 & 0.05613 & 0.05248 & 0.05398 & 0.0580 & 0.0429 & 0.0430 \\
6  & 0.00038 & 0.00144 & 0.00108 & 0.05064 & 0.05204 & 0.04995 & 0.0394 & 0.0419 & 0.0394 \\
8  & 0.00687 & 0.00220 & 0.00050 & 0.04471 & 0.04724 & 0.03052 & 0.0324 & 0.0367 & 0.0171 \\
10 & 0.00103 & 0.00018 & 0.00021 & 0.03241 & 0.03178 & 0.03040 & 0.0208 & 0.0181 & 0.0169 \\
\hline
\end{tabular}
}
\end{table}

Finally, additional experiments were conducted to evaluate performance when modifying the selection threshold defined in Equation \ref{eq_12}. The results are summarized in Table \ref{tab:results_3}, where the threshold was varied from 95\% to 75\%, 55\%, and 45\%. For these tests, the parameters from T15 were adopted. As observed, lowering the threshold to 75\% improves performance across all six metrics ($\approx 10\%$ due to the increased diversity of the training set. A further reduction to 55\% yields a slight additional improvement. However, when the threshold is decreased to 45\%, performance drops, as introducing a bias in the estimation degrades the quality of pseudo-labels produced by the co-training method. This illustrates a fundamental trade-off in semi-supervised learning: while incorporating unlabeled data improves performance by increasing diversity, maintaining a minimum quality threshold is essential to avoid the propagation of noisy labels and systematic biases.

\begin{table}[!ht]
\centering
\caption{Results of the experimental evaluation}
\adjustbox{max width=\columnwidth}{%
\begin{tabular}{|c|c|c|c|}
\hline
\multicolumn{4}{|c|}{\textbf{Tests with 10\% of labeled data}}\\
\hline
\multirow{2}{*}{\textbf{Metric}} & T16 & T17 & T18 \\
& \textbf{THOLD = 75\%} & \textbf{THOLD = 55\%} & \textbf{THOLD = 45\%} \\
\hline
\textbf{MAE} & 0.04141 $\pm$ 0.00012 & \textcolor{red}{0.04117 $\pm$ 0.00030} & 0.04162 $\pm$ 0.00009 \\
\hline
\textbf{$R^{2}$} & 0.90243 $\pm$ 0.00130 & \textcolor{red}{0.90369 $\pm$ 0.00228} & 0.90156 $\pm$ 0.00070 \\
\hline
\textbf{MSE} & 0.00449 $\pm$ 0.00006 & \textcolor{red}{0.00443 $\pm$ 0.00011} & 0.00453 $\pm$ 0.00003 \\
\hline
\textbf{$\Delta z$} & \textcolor{red}{0.00008 $\pm$ 0.00011} & 0.00030 $\pm$ 0.00040 & 0.00042 $\pm$ 0.00006 \\
\hline
\textbf{$\sigma_{zp}$} & 0.03040 $\pm$ 0.00009 & \textcolor{red}{0.03033 $\pm$ 0.00016} & 0.03060 $\pm$ 0.00008 \\
\hline
\textbf{Out\_Frac} & 0.01675 $\pm$ 0.00044 & \textcolor{red}{0.01644 $\pm$ 0.00078} & 0.01719 $\pm$ 0.00026 \\
\hline
\end{tabular}%
}
\label{tab:results_3}
\end{table}

For test 17, the evolution of metrics $\Delta z$, $\sigma_{68}$, and outlier fraction was analyzed as a function of redshift to understand Co-SOM performance across the redshift range better. The bias ($\Delta z$) remains close to zero at low and intermediate redshifts, indicating accurate photometric redshift estimates in these ranges. At redshifts $z \gtrsim 0.6$, the bias ($\Delta z$) trends negative, accompanied by increased scatter ($\sigma_{68}$) and higher outlier fractions, reflecting larger photometric uncertainties and the limited availability of spectroscopic data. These results demonstrate that Co-SOM maintains robust performance across most of the redshift range, with degradation only at the highest redshifts where spectroscopic training data are limited. This analysis emphasizes the importance of evaluating photometric redshift methods not only globally but also as a function of redshift, to identify potential limitations and guide future improvements.

\begin{table}[h!]
\centering
\caption{Evolution of metrics ($\Delta z$, $\sigma_{68}$, and outlier fraction) as a function of redshift for test 17.}
\adjustbox{max width=\columnwidth}{%
\begin{tabular}{|c|c|c|c|}
\hline
Redshift bin & $\Delta z$ & $\sigma_{68}$ & Outlier fraction \\ 
\hline
$0.0 - 0.1$ & 0.0151 & 0.0241 & 0.0123 \\
$0.1 - 0.2$ & -0.0012 & 0.0237 & 0.0120 \\
$0.2 - 0.3$ & 0.0099 & 0.0288 & 0.0490 \\
$0.3 - 0.4$ & 0.0118 & 0.0320 & 0.0381 \\
$0.4 - 0.5$ & 0.0104 & 0.0322 & 0.0118 \\
$0.5 - 0.6$ & -0.0037 & 0.0294 & 0.0096 \\
$0.6 - 0.7$ & -0.0250 & 0.0502 & 0.0045 \\
$0.7 - 0.8$ & -0.0403 & 0.0698 & 0.0020 \\
$0.8 - 0.9$ & -0.0568 & 0.0691 & 0.0004 \\
$0.9 - 1.0$ & -0.1024 & 0.1175 & 0.0 \\
\hline
\end{tabular}%
}
\label{tab:metrics_vs_z}
\end{table}

\section{Conclusions}

With the upcoming generation of large galaxy surveys over the next decade, such as the Legacy Survey of Space and Time (LSST), millions of galaxies will require high-precision redshift measurements and multi-band photometry. However, the current scarcity of spectroscopic data poses a significant challenge, necessitating the development of computational methods that leverage the limited spectroscopic datasets to derive precise redshift estimates. In this context, semi-supervised learning approaches, such as Co-SOM, will be highly valuable as they utilize both labeled and unlabeled data. In this work, we explore different parameter configurations of the map, co-training strategies, and training regions, modifying the quantity of labeled data used during training.

Each experiment was evaluated using six distinct metrics that capture different aspects of redshift estimation during the map training process. A structured methodology was employed to ensure reproducibility, including data selection via the query described in Appendix A, pre-processing stages with their corresponding parameters, construction of the training sample through stratified sampling, and evaluation of the SOM and co-training parameters. Additionally, the SOM’s topological properties were utilized to inform the co-training selection method. In addition, the impact of progressively incorporating more labeled data and varying the Minimum Quality Criterion was analyzed, finding that a broader threshold during training led to improved redshift estimation. Runtimes were monitored across all test sets, showing that Co-SOM can be trained efficiently on the described server, with training times increasing moderately as the map size increased. At the same time, larger fractions of spectroscopic data reduced the runtime per object.

It is worth noting that, at the current stage, Co-SOM produces point estimates of photometric redshifts rather than full posterior PDFs like SOM-based approaches used by KiDS and DES. However, Co-SOM using only $1\%$ spectroscopic redshifts ($\approx20k$ galaxies), it achieves competitive bias levels, approaching those required for LSST. In future work, we will attempt to estimate the $N(z)$ distributions needed for weak lensing or large-scale structure analyses.

Additionally, Co-SOM is sensitive to galaxy photometric properties, as are other photometric redshift methods, and its performance may be affected by survey-specific systematics such as variations in depth, extinction, or seeing. Nevertheless, the bias mitigation steps, the use of multiple maps, a well-characterized sample, and additional strategies such as enhanced data augmentation, mock simulated datasets, robust outlier handling, and consideration of additional galaxy properties, including the angular effective radius, could help control estimation uncertainties and break potential systematic degeneracies.

Our SDSS sample was restricted to non-inclined and well-resolved galaxies, as well as to the redshift range $0.005 < z < 1.0$, whereas surveys such as Rubin will rely on higher-redshift sources. These criteria do not limit the method; the approach remains general and applicable to broader galaxy samples, including those relevant for Rubin and other forthcoming large-scale surveys.

Future work will extend the analysis of high-$Z$ performance by training Co-SOM on deeper and wider surveys (e.g., COSMOS, HSC-SSP, DEEP2) and on realistic mock simulated datasets. We will also explore adaptive weighting schemes and parameters that leverage the topological properties of the SOM to mitigate the propagation of noisy pseudo-labels and enhance redshift estimation in sparsely labeled or low-S/N regimes.

\section*{Acknowledgements}
O.V., E.M., and A.C. acknowledge support from the SECIHTI grant CF-2023-G-1052 and DGAPA UNAM grants  AG102123, IG101725, and IG101222. Some of the calculations in this work were carried out on the HPC clusters Atocatl and Tochtli at LAMOD-UNAM. LAMOD is a collaborative project of IA, ICN, IQ Institutes, and DGTIC at UNAM.

\section*{Declaration of Competing Interest}

The authors declare that they have no known competing financial interests or personal relationships that could have appeared to influence the work reported in this paper.

\appendix

\section{MySQL data query}
The dataset used in this work was derived from the SDSS CasJobs Web portal through MySQL queries, as outlined below. The data were downloaded in stages by applying redshift (Z) cuts of 0.1, due to the substantial number of galaxies within the sample. This resulted in a total of 2,155,735 galaxies after applying the Z cuts, which are further described in the "Data" section.

\begin{verbatim}
SELECT  pa.u, pa.g, pa.r, pa.i, pa.z, 
spa.z, pa.err_u, pa.err_g, pa.err_r, pa.err_i,
pa.err_z, spa.zErr
INTO mydb.Table_1 
FROM PhotoObjAll AS pa
JOIN SpecPhotoAll AS spa
  ON pa.ObjID = spa.ObjID
WHERE
  spa.zWarning = 0
  AND spa.class = 'GALAXY'
  AND pa.u > 0
  AND pa.g > 0
  AND pa.r > 0
  AND pa.i > 0
  AND pa.z > 0
  AND pa.deVAB_r > 0.4
  AND (pa.petroRad_r / pa.psffwhm_r) > 1
  AND spa.z BETWEEN 0.005 AND 0.1
\end{verbatim}

In this query, the specified parameters filter the data for valid galaxy objects with well-defined photometric and spectroscopic properties, ensuring that only galaxies with accurate redshift values ($0.005 \le z \le 0.1$), high signal-to-noise ratios (as indicated by photometric quality thresholds), and reliable measurements for structural parameters are included in the final dataset. The inclusion of redshift constraints aims to select galaxies within the desired range for subsequent analysis while minimizing contamination from non-galaxy objects or unreliable measurements.
\bibliographystyle{elsarticle-harv} 
\bibliography{bibliography}






\end{document}